\documentclass[10pt]{article}
\usepackage[OE]{express}

\begin{document}
\title{A High-Mechanical Bandwidth Fabry-Perot Fiber Cavity}

\author{Erika Janitz,\authormark{1*} Maximilian Ruf,\authormark{1} Yannik Fontana,\authormark{1} Jack Sankey,\authormark{1} and Lilian Childress\authormark{1}}

\address{\authormark{1}Physics Department, McGill University, 3600 Rue University, Montreal QC Canada, H3A 2T8.}

\email{\authormark{*}erika.janitz@mail.mcgill.ca} 



\begin{abstract}
Fiber-based optical microcavities exhibit high quality factor and low mode volume resonances that make them attractive for coupling light to individual atoms or other microscopic systems. Moreover, their low mass should lead to excellent mechanical response up to high frequencies, opening the possibility for high bandwidth stabilization of the cavity length. Here, we demonstrate a locking bandwidth of 44 kHz achieved using a simple, compact design that exploits these properties. Owing to the simplicity of fiber feedthroughs and lack of free-space alignment, this design is inherently compatible with vacuum and cryogenic environments. We measure  the transfer function of the feedback circuit (closed-loop) and the cavity mount itself (open-loop), which, combined with simulations of the mechanical response of our device, provide insight into underlying limitations of the design as well as further improvements that can be made.
\end{abstract}


\section{Introduction}

High finesse Fabry-Perot cavities have extensive applications in spectroscopy \cite{berden2000cavity,srinivasan2007linear} and precision measurement \cite{udem2002optical,hinkley2013atomic}, as well as fundamental research in quantum optics \cite{leibfried2003quantum,birnbaum2005photon}. In many situations,  the cavity must be locked to a frequency reference (for example, an atomic transition) to compensate for external disturbances and maintain a specific resonant frequency \cite{mabuchi2002cavity,birnbaum2005photon,keller2004continuous,steiner_ion}. Such locking is typically accomplished by monitoring the transmission or reflection of light at the reference frequency, and feeding the signal back to a mechanical transducer that adjusts a mirror to stabilize the cavity length. The bandwidth of the lock (here defined as the frequency below which noise is suppressed) represents a key figure of merit, and determines the maximal noise suppression that can be achieved at low frequencies. The mechanical response of the mirrors and mirror mount typically limit the bandwidth, requiring careful engineering to suppress low-frequency vibrational modes.

Fiber-based micro-mirrors\cite{Hunger2010_NJP} offer a promising technology for creating tunable, high-finesse micro-cavities. These cavities can achieve very small mode waists, which are advantageous for cavity quantum electrodynamics applications; so far, they have been coupled to atoms\cite{gehr_atom_hyperfine,haas_atom_mentangled,colombe_bose}, ions\cite{steiner_ion,balance2017ion}, optomechanical systems \cite{kashkanova2017superfluid,flowers2012fiber}, molecules\cite{toninelli2010scanning}, and crystalline defect centers\cite{albrecht2013NV,benedikter2016SiV,riedel2017deterministic}. Moreover, the light weight of the fiber mirror suggests that it should be possible to achieve a high bandwidth feedback loop for length stabilization of such a cavity.

Previous work on locking fiber cavities has demonstrated mechanical feedback with bandwidths of only 1-3 kHz\cite{gallego2016, brachmann2016} (we note that for the 1 kHz case the first limiting mechanical resonance occurred at 25 kHz). Adding photothermal stabilization can further improve noise suppression for frequencies up to 500 kHz\cite{gallego2016, brachmann2016}. Such ``self-stable" operation is achieved via intra-cavity heating of the mirror coatings by an incident laser; disturbances that change the length of the cavity affect the intra-cavity power, which in turn induces thermal expansion that stabilizes the effective length. This method of thermal stabilization comes at the cost of high intra-cavity power on the order of 1-10 Watts\cite{gallego2016, brachmann2016}, presenting a challenge for cryogenic operation or for coupling cavities to solid-state systems where non-resonant absorption of the locking light cannot be neglected. 

In contrast, higher locking bandwidths can also be achieved using careful mechanical and electrical engineering of the cavity mount and feedback circuit. An optimized design for a macroscopic Fabry-Perot cavity composed of two small free-space mirrors achieved a bandwidth of up to 180 kHz\cite{briles2010}. However, this result relied on the damping properties of lead inside the mirror mount to reduce the impact of low-frequency mechanical resonances on the feedback circuit, which limits its function to non-cryogenic applications.

Here, we investigate the bandwidth attainable when electronically feeding back to a piezo-mounted fiber mirror. Our approach does not rely on specific material properties or intra-cavity heating; instead, we take advantage of the intrinsic high-frequency response available with a lightweight fiber mirror. We measure the full transfer function for the feedback circuit, and find that a locking bandwidth of 44 kHz is readily obtained. 
With a combination of direct measurements of the system's transfer function and finite-element simulations, we identify limiting features in the mechanical response associated with resonances in the mount, fiber, and epoxy, and provide an additional set of design considerations. 

\section{Experimental Setup}
\begin{figure}
\centering\includegraphics[scale=1]{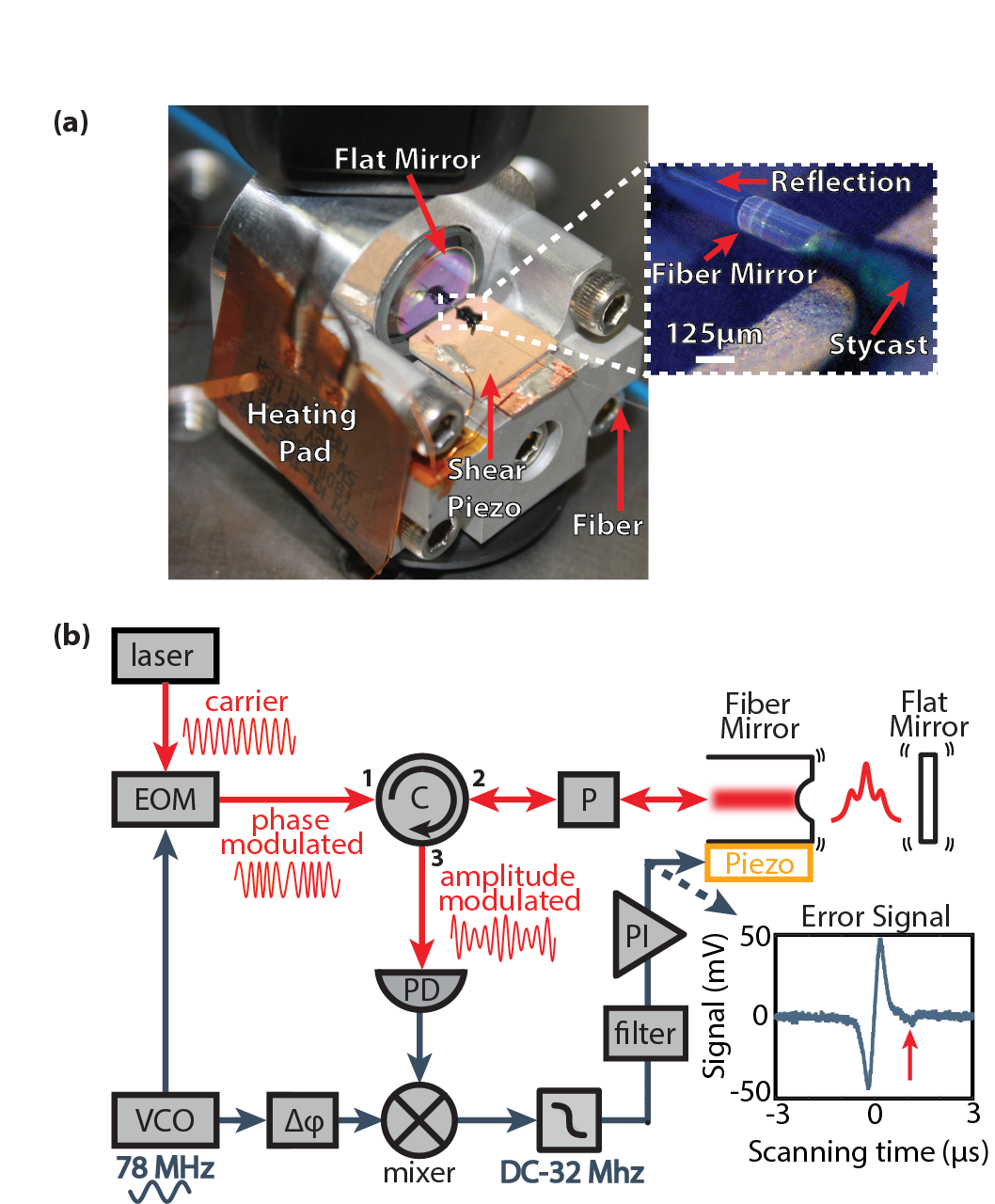}
\caption{\label{fig1} a) An image of the assembled fiber cavity device (inset: a microscope camera image showing the fiber mirror glued to the shear piezo, where the reflection can be seen in the flat mirror). b) A schematic showing the Pound-Drever-Hall locking circuit and a sample error signal measured for our cavity. The low amplitude nonideality visible to the right of the error signal (indicated with a red arrow) corresponds to the incompletely suppressed orthogonal linear polarization mode. EOM: electro-optic modulator; VCO: voltage controlled oscillator; PD: photodiode; C: circulator; filter: interchangeable analog filter circuit; P: polarization control; PI: proportional-integral amplifier.}
\end{figure}

\subsection{Device Design and Construction}
Our cavity is formed by a macroscopic flat mirror and a microscopic spherical mirror fabricated on the tip of a single mode optical fiber (Fig.~\ref{fig1}a), using a $\text{CO}_2$ laser ablation process\cite{Hunger2010_NJP,janitz2015}. Both mirror substrates are coated with a high reflectivity dielectric mirror (LASEROPTIK) with the reflection band centered at 1550 nm. This coating exhibits a finesse of $\mathcal{F}$=21000$\pm$2000, measured after annealing for 5 hours at $300^\circ$C under atmospheric conditions\cite{Brandstatter2013}.

The device is composed of two aluminum pieces onto which each mirror is mounted. The assembled device has dimensions of 28 mm in all directions. This aspect ratio was chosen heuristically to maximize the frequencies of the structure's normal modes, thereby minimizing the coupling between ambient (or driven) vibrations and the cavity length. The upright aluminum piece serves as a mount for the flat mirror, and (in this case) the mirror is glued in place using Stycast 2850.

On a second aluminum piece, a shear piezoelectric actuator (Noliac CSAP03) supports the fiber mirror and controls the length of the cavity. This piezo was chosen for its high unloaded resonance frequency of 1.75 MHz, which ultimately limits the theoretically achievable locking bandwidth. The bottom electrode is connected to a large DC voltage to coarsely tune the cavity length, while the top electrode is connected to a fast signal for feedback (maximum $\pm$10 V). The combined voltages we can achieve limit our travel range to $\approx600$ nm at room temperature, less than a free spectral range of our cavity ($\lambda/2=775$ nm). We therefore employ a heating pad and thermocouple to control the temperature of the aluminum, and tune a cavity resonance within range of the piezo. For the case of a cryogenic environment, piezo travel should decrease by a factor of $\sim$5, requiring a combination of improved fabrication tolerances, higher drive voltages, and/or longer-travel piezo elements.

To mount the actuator, an alumina plate is first glued (Stycast 2850) to the second aluminum piece for electrical insulation. A piece of copper slightly longer than the piezo is then glued to the alumina, where the exposed copper is used for electrical contact to the bottom electrode. The piezo is attached to the copper sheet using silver epoxy (Epotek H20E), and kapton-coated copper wires are similarly glued to the copper sheet and top electrode. The two aluminum pieces are then screwed together.

Finally, the fiber mirror is pre-aligned above the actuator and glued in place with Stycast 2850 (the fiber is further aligned while submerged in the epoxy to maximize the contrast of the cavity reflection dip for a fundamental mode). The system is left to cure for 24 hours under ambient conditions. 

The assembled device is then clamped to an optical table between two viton O-rings (visible above and below the device in Fig.~\ref{fig1}a) to help isolate the system from high-frequency noise transmitted through the table.

\subsection{Pound-Drever-Hall Locking Circuit}
To lock the cavity length to our laser frequency, we employ active feedback via the Pound-Drever-Hall locking scheme \cite{drever1983,black_pdh}. Briefly, the incident laser frequency is modulated and the reflected light is demodulated to produce an error signal that depends  linearly on detuning from the cavity resonance. This signal is filtered, amplified, and fed back to the shear piezo under the fiber mirror.

The circuit used for generating an error signal is shown in Fig.~$\ref{fig1}$b, with optical (electrical) signal paths indicated in red (blue). The laser (Koheras Adjustik E15) is frequency modulated using an electro-optic phase modulator (EOM), driven with the output of a voltage controlled oscillator (VCO). The VCO is capable of generating output frequencies $f_m$=76-80 MHz, which corresponds to the so-called ``low-modulation regime"\cite{black_pdh} for our cavity, which has a (full) linewidth of 189$\pm$7 MHz. We find it is important to use only APC terminated fibers as well as isolators in our optical circuit to prevent back-reflections and standing waves that lead to residual amplitude modulation \cite{whittaker1985}. The modulated beam passes through a circulator (C) and couples to the cavity through the fiber mirror. The degeneracy of the two linear polarization states of the fundamental cavity mode is lifted by $\sim$100 MHz, likely due to a combination of fiber mirror ellipticity and birefringence\cite{uphoff2015frequency}. The polarization is adjusted to select only one of these modes using a fiber polarizer (P), which we correct on an hourly basis as the strain in the fibers changes with lab temperature (typical temperature excursions $<1$ $^\circ$C).

The reflected power is measured on a high bandwidth photodetector (bandwidth 150 MHz), and the output is demodulated by mixing with the VCO signal. The mixer output is sent through a 32 MHz low pass filter to remove the $f_m$ and $2f_m$ oscillating terms, and the relative phase between the photodiode and mixing signal ($\phi$) is adjusted by tuning the VCO frequency to obtain the desired error signal \cite{black_pdh}, similar to what is shown in the inset of Fig.~\ref{fig1}b. The residual signal from the incompletely suppressed orthogonal polarization mode is also visible to the right of the main features (indicated by the red arrow). This error signal is then sent through a combination of filters and a servo controller consisting of a ``proportional-integral" (PI) amplifier (Newfocus LB1005) before finally being applied to the shear piezo. As discussed below, the filters and  controller allow us to maximize the feedback gain at low frequencies for a given lock bandwidth.  

\section{Closed-Loop Measurements}
\subsection{Measurement Block Diagram}
We can quickly gain insight into the limitations of this feedback system and the mechanical response of the mount by injecting a disturbance into the locked circuit and measuring its response, as described in Ref. \cite{reinhardt2016}. As discussed below, this provides an estimate of the closed-loop transfer function.

A block diagram representing our feedback system is shown in Fig.~\ref{fig2}a, where all circuit element transfer functions and signals are complex functions of frequency. The error signal ($a$) is connected to the positive input of our servo controller, which also has an inverting input ($b$). The controller has an error monitor port that produces a voltage $e = T(a-b)$, where $T$ is the associated transfer function. The controller itself has a PI transfer function $P$, such that its output voltage is $u = P(a-b)$. The output $u$ then drives the system ($G$), which represents transduction between applied voltage and change in cavity length. Practically, this comprises a 200 $\Omega$ resistor in series with the piezo actuator supporting the fiber mirror. These two elements form a low-pass filter with a cut-off frequency $\approx240$ kHz. We found that it was not possible to lock the system without a resistor, as the low-pass filter behaviour is essential to suppress high-frequency perturbations. Consequently, the resistor is included in the system transfer function $G$ for all closed-loop measurements. The piezo converts the output voltage into shear displacement (with a harmonic-oscillator-like transfer function), and all mechanical noise experienced by the cavity is modeled by the addition of a driving term $d$. The deviation of the length of the cavity from resonance is measured by the cavity mirrors, photodiode, and mixer circuit, which together have transfer function $-M$. $M$ represents the cavity light's dynamical response to mechanical motion (essentially a low-pass filter with time constant equal to the cavity's ringdown time\cite{reinhardt2016,rakhmanov2002dynamic}), along with the transfer functions of the fiber components, photodiode, cables, mixer, and 32 MHz low-pass filter. The sign of $M$ is determined by the phase of the local oscillator, and is negative in our case. As discussed below, we also have the freedom to introduce an additional filter $F$ prior to the PI controller, where the output is the error signal $a$ that is fed back to the servo controller.

To probe the frequency response of different elements in the loop, we can apply a known perturbation $b$ on the inverting input of the servo controller and observe the closed-loop response. Solving for the measurable quantity $e$ in terms of the inputs $b$ and $d$ yields:
\begin{align}
e &= \frac{-T}{1+PGMF}b-\frac{-TMF}{1+PGMF}d \label{e_meas}.
\end{align}
All of the perturbations acting on the locking system are scaled by a term proportional to $1/(1+PGMF)$. Thus, when the magnitude of $PGMF$ is large, the effect of these disturbances is minimized and the error signal tends to 0. Conversely, if $PGMF$ approaches the value -1 at some frequency, we enter a situation of positive feedback where the signal on the inputs is amplified \cite{bechhoefer2005}. Using a dual-phase lock-in amplifier (Zurich Instruments HF2LI) to supply the perturbation $b$ and to measure $e$, the noise term $d$ (which is uncorrelated with the lock-in's output $b$) can be eliminated, and it is straightforward to extract the circuit transfer function:
\begin{align}
	PGMF&=-\frac{Tb}{e}-1 \label{closedloop}.
\end{align}
Here the only unknown is the error monitor port transfer function $T$, which can be measured independently (i.e. while unlocked). We hereafter refer to $PGMF$ as the transfer function of the circuit.

\subsection{Closed-loop Measurements}
\begin{figure}
\centering\includegraphics[scale=1]{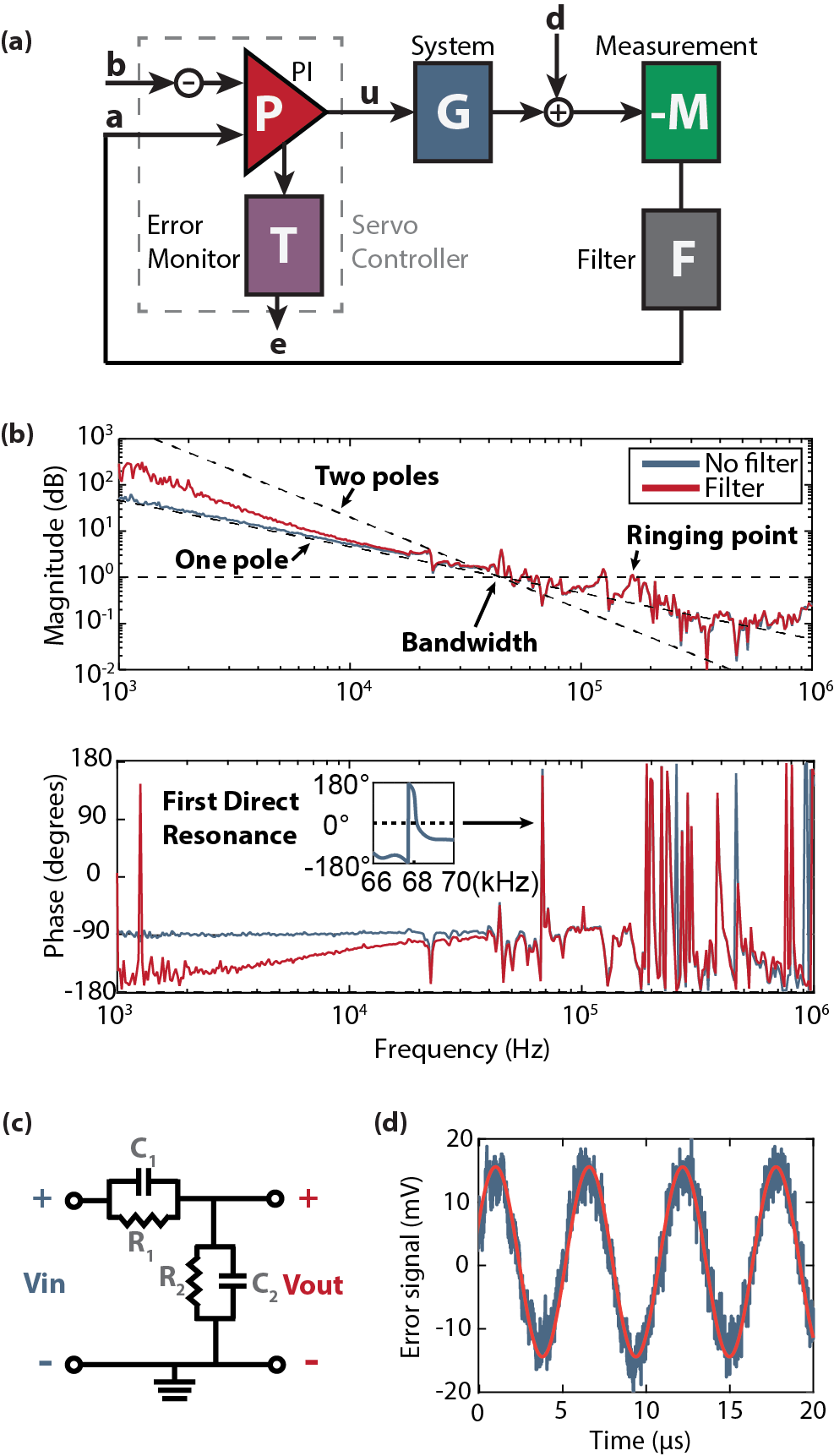}
\caption{\label{fig2} a) A block diagram illustrating the feedback loop. All circuit elements and signals are analyzed as a function of frequency.  b) Circuit transfer functions extracted for the two different filter configurations (as described in the text). The circuit bandwidth, first ringing point, and first direct resonance are indicated with arrows (inset: high resolution plot of the measured phase response about the first direct resonance). The low frequency gains given one and two poles of roll-off from the bandwidth frequency are indicated with dashed lines.  c) A circuit diagram for the electronic filter $F$. d) A time trace of the error signal where the proportional gain has been increased to cause ringing at 179 kHz (fit overlaid in red).}
\end{figure}
The extracted circuit transfer functions can be seen in Fig.~\ref{fig2}b for  the ``no filter" ($F=1$) case, and for the case where $F$ is an analog electronic filter (schematic shown in Fig.~\ref{fig2}c). The filter increases the noise suppression of the lock at low-frequencies by adding a second pole of roll-off to $PGMF$ from $\approx$ 300 Hz - 5 kHz, with the transfer function:
\begin{align}
F=\frac{R_2(C_1\, R_1\, \omega-i)}{(C_1+C_2)R_1\,R_2\, \omega-i(R_1+R_2)}
\end{align} 
where $C_1=330$ pF, $C_2=4.7$ nF, $R_1$=100 k$\Omega$, and $R_2$=1 M$\Omega$.

For the ``no filter" case, $PGMF$ exhibits one pole of roll-off at low frequencies from the PI controller, which has transfer function $P=K(1-i\frac{\omega_{PI}}{\omega})$, where $K$ is an overall scaling factor, and $\omega_{PI}$ is the ``PI corner" frequency of $\omega_{PI}/2\pi$=100 kHz. The inclusion of the additional electronic filter allows us to attain a higher low frequency gain ($PGMF=272\pm 40$ at 1 kHz), but does not affect the transfer function considerably at higher frequencies. 

Between 20-68 kHz some small resonances appear that are mainly related to motion of the macroscopic mirror, deformation of the aluminum jig, and bending resonances of the fiber tip, as discussed in detail in Section \ref{sec3}. We refer to these as ``indirect" resonances since they correspond to small phase excursions of $<\pi$, and could in principle be compensated for with electronics. Starting at 68 kHz, an increasingly dense set of resonances appears, stemming from the mechanical modes of the fiber mount and the clamped fiber. We refer to these resonances as ``direct", since they behave similarly to a directly driven harmonic oscillator, exhibiting a phase decrease on the order of $\pi$ when passing through resonance.
 \subsubsection{Circuit Bandwidth} We define the bandwidth of our circuit as the frequency range over which $|PGMF|>1$ and $\text{arg}(PGMF)>-\pi$; if the limiting conditions occur at the same frequency, it corresponds to the ringing condition of our system ($PGMF=-1$). One can see that for both plots in Fig.~\ref{fig2}b, the first instance of $|PGMF|=1$ occurs at 44 kHz, while $\text{arg}(PGMF)>-\pi$ for frequencies up to 68 kHz (the first direct resonance). We therefore conservatively define the bandwidth of our system as 44 kHz for both measurements (the apparent $-\pi$-phase crossing point at 1.5 kHz in the ``filter" plot is a result of measurement noise).
\subsubsection{Ringing frequency}
For many systems, one would expect the amplitude and phase response to decrease monotonically with frequency after the first direct resonance (as, for example, in the case of a simple harmonic oscillator). In this case, the proportional gain provided by the PI controller could be increased until $|PGMF|=1$ at the lowest frequency where $\text{arg}(PGMF)=-\pi$, and the system would ring at this frequency. The electro-mechanical modes of our system are not so simply modeled, and result in many high frequency resonances that exhibit an increase in magnitude and phase. Consequently, the bandwidth, first zero-phase crossing frequency, and first ringing frequency are all different. Fig.~$\ref{fig2}$d shows a time trace of the the locked error signal where the proportional gain has been increased to hit the first ringing point  at 179 kHz.
\section{System Transfer Functions}
\label{sec3}
\begin{figure}
\centering\includegraphics[scale=1]{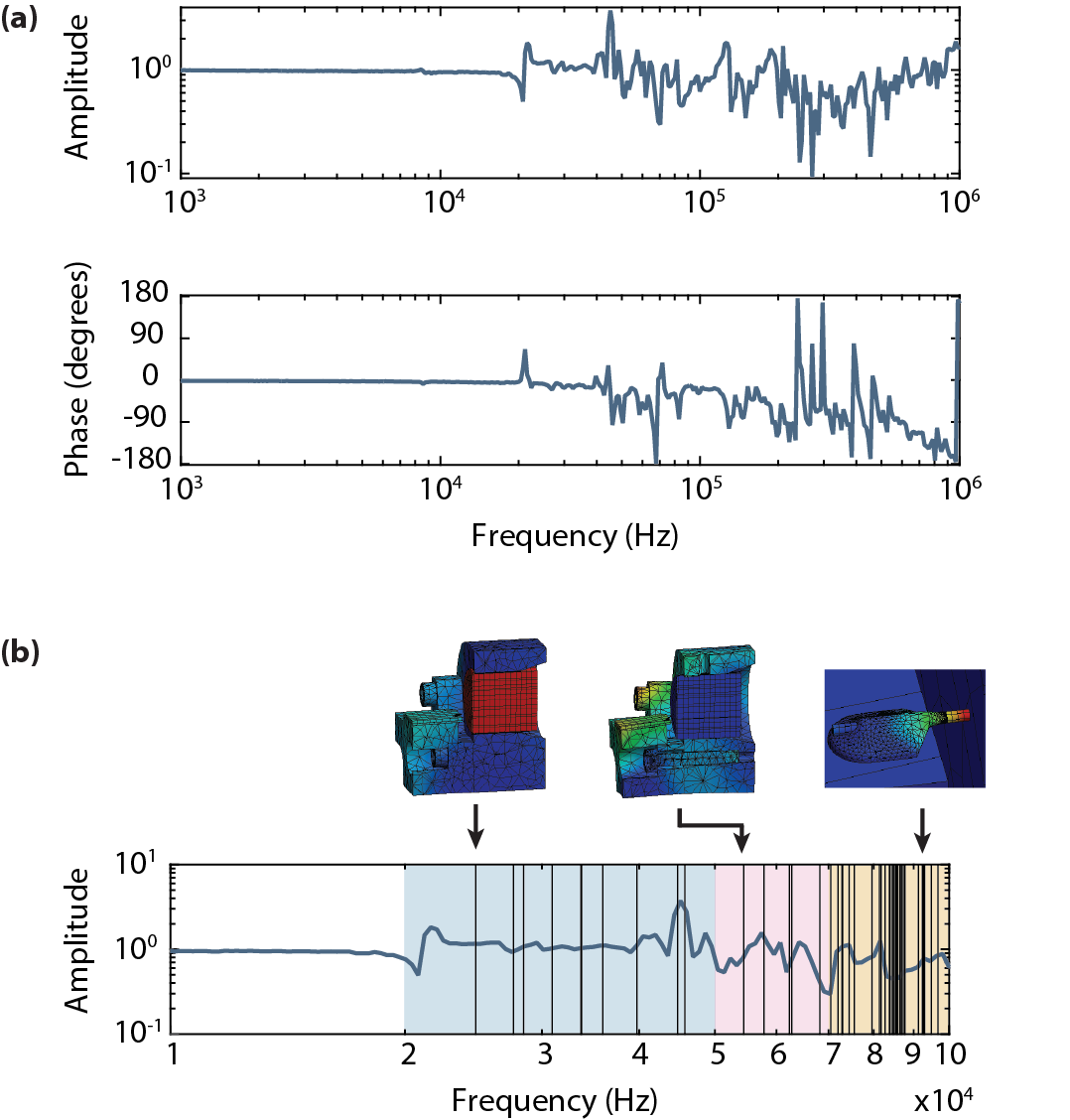}
\caption{\label{fig3} a) The system transfer function normalized to the low frequency magntitude. b) The amplitude of the measured system transfer function overlaid with a range of simulated mechanical resonant frequencies (black vertical lines). The low, medium, and high frequency regions as discussed in Section \ref{sec3} are indicated by the blue, red, and orange shaded regions respectively. The simulated mechanical displacement for three typical resonances from the different regions are shown above (images used courtesy of ANSYS, Inc.), where the first mode corresponds to slipping motion of the flat mirror, the second mode corresponds to flexing of the jig under the piezo, and the third mode corresponds to motion of the fiber along the optical axis. These representations are exaggerated with respect to real displacement for clarity, with elements in red (blue) being subjected to larger (smaller) displacement for a given mode.}
\end{figure}
As discussed in the previous section, the mechanical response of our cavity has a complicated structure. It  is therefore of interest to isolate the system transfer function $G$ to determine limiting factors in the design with regard to the locking bandwidth. 

We measure the response of the cavity mount ($G$) by coupling a 1310 nm laser through the fiber mirror where the dielectric mirror coatings are low finesse (with power reflection coefficient R$\approx$45$\%$). A high voltage DC signal is first applied to the bottom electrode of the piezo to select a cavity length offset corresponding to high measurement sensitivity (approximately the point of highest slope on the reflection fringe). An AC drive voltage from a lock-in amplifier is then applied to the top electrode while recording the modulation in reflected light, thereby probing the mechanical response of the system at different frequencies. The resulting system transfer function is shown in Fig.~\ref{fig3}a (normalized to the low frequency amplitude). 

We simulated the mechanical modes of the assembled cavity device using the ANSYS \cite{ansys} finite element analysis program to understand the origin of different resonances. The results are overlaid with the measured amplitude response between 10-100 kHz in Fig.~\ref{fig3}b, which yield a decent correspondence given the difficulties in modeling the exact system. Interestingly, the simulation suggests that the mechanical resonances can be divided into three distinct frequency ranges:

\begin{itemize}
\item The low frequency region (20-50 kHz) contains only a few modes and is mainly characterized by the deformation of the jig and the slip/rotation of the macroscopic flat mirror. We attribute the measured resonance near 20 kHz to the movement of the flat mirror in its housing (this mode is calculated to occur at 24.7 kHz in our simulations). These resonant frequencies could be improved (increased) by optimizing the geometry of the mount, reducing the mass of the macroscopic mirror, and changing how the mirror is fixed to the mount (for example, using a flexural clamp similar to a shaft collar).
\item The mid-frequency region (50-70 kHz) is characterized by the low frequency bending resonances of the overhanging fiber tip, as well as the bending/folding of the jig under the piezo. Our simulations suggest that further reducing the length of overhanging fiber (from $L\approx500$ $\mu$m) would increase the frequencies of the clamped fiber modes, where the resonant frequencies scale approximately with $1/L^2$ (up to frequencies where the mechanical modes of the glue structure are excited). 
\item The high-frequency region ($>$70 kHz) is the most interesting with regard to fiber mounting considerations. In this frequency range, the fiber no longer acts as a simple clamped beam and the mechanical modes begin to incorporate motion of both the fiber and the glue bonding it to the piezo. The glue will deform along the fiber axis when driven with shear motion, impacting the cavity length directly. This could be improved by ensuring the fiber is in contact with the piezo while the epoxy cures, and by using less epoxy.

At these high frequencies it is also necessary to consider vibration of the clamping screws, and ``flapping" at unbonded corners for the different layers in the piezo stack.
\end{itemize}

\subsection{Cryogenic Operation}
\begin{figure}
\centering\includegraphics[scale=1]{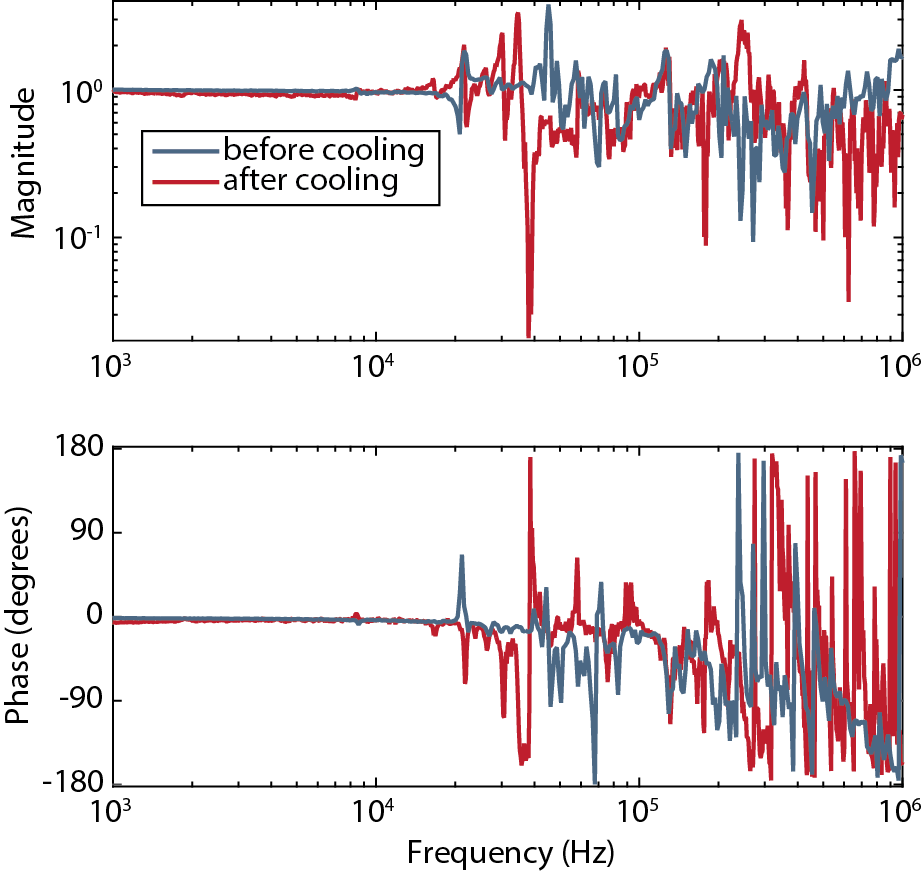}
\caption{\label{fig4}
The system transfer function ($G$) before and after thermal cycling. The fiber mirror used before thermal cycling was replaced to facilitate a longer cavity length (shorter fiber overhang), and the piezo stack is unchanged.}
\end{figure}
Many experiments in quantum optics require moving to cryogenic temperatures where mechanisms for thermal decoherence are suppressed. Although our device has not been expressly designed for cryogenic operation, its materials are in principle cryogenically compatible, and thus we explore the impact of thermal cycling on the system. We clamp our device using a stainless steel strap to the base plate of a Montana Instruments Nanoscale Workstation cryostat. The fiber mirror used for the previous measurements was replaced to facilitate a longer cavity length of $\approx60\ \mu$m that would prevent the fiber from crashing into the flat mirror due to thermal contraction of the mount. We estimate that this increase in cavity length corresponds to reducing the overhanging fiber length by $\sim10\%$. The piezo stack is unchanged to allow for comparison with previous measurements. 

The device survived two thermal cycles from room temperature to 6K. Due to the limited range of our piezo and voltage supplies, we were not able to make measurements at the base temperature. Nevertheless, after thermal cycling, we were able to maintain a cavity lock at 270K with the cryo-cooler running, but the system transfer function G changed considerably (see Fig.~\ref{fig4}), likely due to a loosening of the aluminum screws holding the device together or delamination of the piezo stack. The new strong resonance at 37 kHz set a new ringing point for the locking circuit, and limited the achievable lock bandwidth to 3 kHz. The device broke at the epoxy connection between the alumina plate and aluminum mount as we tried to add Belleville washers to the screws, indicating that mount materials with a closer thermal expansion match to alumina (stainless steel, titanium) may prove advantageous for long-term cryogenic operation.

\section{Conclusion}
Fiber-based micro-mirrors offer a new platform for applications of high finesse Fabry-Perot cavities. We showed that a simple electronic feedback circuit can take advantage of the intrinsic high mechanical resonance frequency of a fiber mirror, achieving a lock bandwidth of 44 kHz. We also modeled the system, and our results suggest that the lock bandwidth may be further improved by minimizing the length of overhanging fiber from the piezo, and reducing the thickness of the epoxy holding the fiber. The resonance frequencies of the mount can be further increased by optimizing the geometry of the mount and, in particular, reducing the size (mass) of the flat mirror. Since our locking approach does not require strong intra-cavity laser power or specialized material damping properties, it should be extendable to cryogenic applications, or to systems requiring minimal perturbations to the cavity's light field and/or weak probe beams. Notably, even though our device is not expressly designed for cryogenic operation, we were still able to lock the cavity, clamped to the cold plate, with our closed-cycle cryostat running. This illustrates the potential for fiber-cavity systems to operate in noisy environments, extending the range of applications for high finesse cavity measurements.

\section*{Funding}
NSERC Discovery  (435554-2013 and 418459-12); CRC (950-229003 and 235060); CFI (229003, 228130 and 33488); FRQNT (NC-172619); Alfred P. Sloan Foundation (BR2013-088); Centre for the Physics of Materials at McGill; Institute Transciplinaire d'Information Quantique; Y.F. acknowledges support by a Swiss National Foundation Early Postdoc Mobility Fellowship.

\section*{Acknowledgments}
We thank Christoph Reinhardt, Tina Muller, and Yoichi Miyahara for helpful discussions.


\end{document}